\documentclass{aa} 

\usepackage{graphicx}
\usepackage{txfonts}
\usepackage{hyperref}
\usepackage{xcolor}

\begin{document}

\title{The X-ray luminosity function of stars observed with SRG/eROSITA}

\author{Nicola Locatelli,
  \inst{1}\fnmsep\thanks{email: nicola.locatelli@inaf.it}
  Gabriele Ponti\inst{1,2,3}, Enza Magaudda\inst{4}, Beate Stelzer\inst{4}
  }

\institute{
 INAF-Osservatorio Astronomico di Brera, Via E. Bianchi 46, I-23807 Merate (LC), Italy
 \and
 Max-Planck-Institut f{\"u}r extraterrestrische Physik, Gießenbachstraße 1, 85748 Garching bei M\"unchen, Germany
 \and Como Lake Center for Astrophysics (CLAP), DiSAT, Università degli Studi dell’Insubria, via Valleggio 11, 22100 Como, Italy
 \and
 Institut f\"ur Astronomie \& Astrophysik, Eberhard Karls Universit\"at T\"ubingen, Sand 1, 72076 Tübingen, Germany
 }

\date{Received 15 July 2025 / Accepted 24 September 2025}

\abstract{
Background diffuse X-ray emission is contributed in large part by the emission of point sources not individually resolved. While this is established since decades for the contribution of quasars to the diffuse emission above 1 keV energies, the possible contribution of undetected stars in the Milky Way to the softer band (0.5-1 keV) background emission is still poorly constrained. The overall unresolved X-ray flux from the stars is the product between the stellar spatial distribution in the Milky Way and the underlying X-ray luminosity function (XLF).
In this work, we build the XLF of the stars and study its structure with respect to a set of main-sequence spectral types (F, G, K, M) and evolutionary stages (giants and white dwarfs).
We build the XLF in volume-limited subsamples of increasing maximum distance and corresponding minimum X-ray luminosity using the HamStar catalog of X-ray emitting stars detected in the Western Galactic hemisphere with the extended ROentgen Survey with an Imaging Telescope Array (eROSITA) on board the Spectrum-Roentgen-Gamma (SRG) space observatory. We build an extinction-corrected Gaia color-magnitude diagram and decompose the XLF into the relative contribution of different groups of stars.
We provide an empirical polynomial description of the XLF for the total sample and for the different stellar subgroups that can be used to estimate the unresolved contribution from the stars to the soft X-ray background of the Milky Way in a flux-limited survey.
}


\titlerunning{X-ray luminosity function of stars}
\authorrunning{Locatelli et al.}
\maketitle
%

\section{Introduction}

Knowledge of the number density of a given class of astrophysical sources as a function of their luminosity is an essential tool to build forward models of the population in simulations \cite[e.g.][]{2004ApJ...601L.147B, 2004ApJ...603..690B} and to estimate the potential contribution of point sources of a given type to the unresolved background flux in shallow observations \cite[e.g.][]{2000A&A...353...25M, 2007Ap&SS.309...35S, 2009Natur.458.1142R}. 

Estimates of the typical luminosity of a population must rely on volume-complete samples, that is, samples that collect information on all the sources within a given distance. 
In practice, real instrumentation characterized by a minimum detectable flux $\rm F_{min}$ is unable to detect the faintest sources (or simply $\rm F<F_{min}$). 
A best-case scenario is the one in which the volume explored with a real instrument does not miss any source because the population within a distance $\rm d_{max}$ is characterized by a physical luminosity $\rm L> F_{min} \cdot 4\pi d_{max}^2$.
However, in most cases, either the volume encompassed by the whole population is too large or the minimum detectable flux $\rm F_{min}$ is too large in order to detect all the sources of the population.
The practical solution which is eventually always implemented in real surveys is to reduce the size of the explored volume or study only the subset of the population more luminous than a minimum luminosity $\rm L_{min}\equiv F_{min} \cdot 4\pi d_{max}^2$.

So far, the only available volume-complete census for M dwarfs and FGK stars in X-rays has been limited to the closest 10-20 pc \citep{2023A&A...676A..14C, 2025A&A...694A..93Z}. 
A limitation of using small volumes is that they may be biased towards a subsample of the overall population. The small volume sample may, in fact, introduce a bias with respect to a given property (e.g. spectral type, age, metallicity), simply given the relatively small number of sources. If we were able to observe the whole Galactic volume, the average properties of the Galaxy (e.g. the luminosity of a given spectral type) would be computed correctly by definition, but this is currently not a real case scenario.

X-ray emission from stars arises from magnetically confined hot plasma in the stellar corona, typically heated by magnetic reconnection \citep{2004A&ARv..12...71G}.
A proxy for the X-ray activity of stars is the X-ray-to-bolometric luminosity ratio, $\rm L_X/L_{bol}$. This quantity is a normalized measure of coronal activity, allowing a comparison of activity levels across stars regardless of size or spectral type. It is especially useful for tracing stellar evolution, activity saturation, and age-related trends.
This is due to the dependence of the X-ray luminosity $\rm L_X$ of a star on its fundamental properties such as stellar radius, effective temperature, bolometric luminosity $\rm L_{bol}$ and mass. Therefore, $\rm L_X/L_{bol}$ shows the level of intrinsic X-ray activity of the source, useful for studying the mechanism that produces the X-rays, for example, by studying the dependence of $\rm L_X/L_{bol}$ with age or rotation
\cite[e.g.][]{2004A&A...417..651S, 2005ApJS..160..390P, 2007AcA....57..149K, 2008MNRAS.390..545D,  2015A&A...578A.129J, 2018MNRAS.479.2351W, 2019A&A...628A..41P, 2020A&A...638A..20M, 2021A&A...649A..96J, 2022A&A...661A..29M, 2022A&A...661A..44S}. 

In this work, we aim to extend the knowledge on the X-ray luminosities of stars to larger volumes by computing the X-ray luminosity function (XLF) for X-ray subsamples defined by $\rm L_X>L_{X,min}$ within a maximum given distance $\rm d_{max}(L_{X,min})$. Effectively, hereby we extend the volume limit to distances much beyond 10 pc.
While also computing $\rm L_X/L_{bol}$, in this work we focus on the XLF of the luminosity $\rm L_X$, as we want to obtain a measurement of how much X-ray emission overall is going to be produced by a given portion of the Milky Way disk of $\sim$kpc size. By building large samples that minimize the sensitivity selection bias, we average over the details of particular stellar populations and types and provide a weighted average luminosity that can be used in order to estimate the overall X-ray background from unresolved stars. The exact computation of this background would require to build a forward model of the Solar Neighborhood that assigns the (un)known correct luminosity function to field stars and to every stellar cluster and association at the corresponding 3D location. However, the knowledge required to build such a forward model, while being of fundamental importance, is nor complete nor available. The approach of this work thus provides a very useful tool to approximate the overall stellar X-ray output from wide regions of the Milky Way by computing it over large samples. In other words, while we do not focus on a specific stellar property, we encompass all possible physical mechanisms that produce X-rays from stars with their observed relative fractions in order to provide a description of the total X-ray output.

\section{Data}

The HamStar catalog \citep{2024A&A...684A.121F} provides information on all point sources identified as coronally active stars detected in the first eROSITA all-sky survey (eRASS1, \citealt{2024A&A...682A..34M}) of the Western Galactic hemisphere. 
The catalog reports fluxes $F_X$ in the soft X-ray band (0.2-2.3 keV).
A relevant information provided for each entry in the catalog is the best-matched \textit{Gaia} DR3 counterpart \citep{2021A&A...649A...1G}, thus the \textit{Gaia} distance and color $B_p-R_p$, G band magnitude $G$. We note that the HamStar magnitudes and colors are not corrected for extinction. We use the available extinction/reddening information ($\rm A_{GSP,g};\, E_{GSP}(B_p-R_p)$) from the \textit{Gaia} GSP-Phot Aeneas library\footnote{\url{https://www.mpia.de/gaia/projects/gsp/aeneas}}. This library uses a Bayesian framework in order to estimate stellar parameters also from low-resolution spectra  \citep{2011MNRAS.411..435B} and is currently being implemented into the Gaia data processing pipeline. We use the available information to compute corrected G-band magnitudes $\rm g_0=g-A_{GSP,g}$ (then converted to the absolute magnitude $\rm G_0$) and colors $\rm (B_p-R_p)_0 = (B_p-R_p) - E_{GSP}(B_p-R_p)$. 
The \textit{Gaia} information thus allows to build a color-magnitude diagram ($G_0;\, (B_p-R_p)_0$) and to divide the sample by spectral types.

In addition, by combining the X-ray flux $\rm F_X$ with the \textit{Gaia} distance we can compute the X-ray luminosity $\rm L_X$ for all the stars from the HamStar catalog. 

\section{Method}

The HamStar catalog is a flux-limited sample of the X-ray emitting stars, an additional selection is thus necessary in order to build a volume-limited sample for computing the XLF. In order to build volume-limited subsamples of HamStar entries, we choose a set of maximum distances $\{\rm d_{max,i}\}$. Within a given (increasing) $\rm d_{max,i}$, we select all the stars with an X-ray luminosity larger than $\rm L_{min,i} = F_{min} * 4\pi d_{max,i}^2$. 
Each subsample is then volume-limited by construction for stars above the $\rm L_{X,min}$ threshold defined by its maximum distance $\rm d_{max,i}$.

As flux limit we use $\rm F_{min} = 5\times 10^{-14}\, erg\, s^{-1}\, cm^{-2}$. This value is the most frequent flux bin in the HamStar catalog and equal to the eRASS1 equatorial detection threshold estimated by \citealt{Predehl2021}. 
This choice is also similar to the value found by \cite{2022A&A...661A..29M} of $\rm F_{min} \sim 3\times 10^{-14}\, erg\, s^{-1}\, cm^{-2}$ for a sample of $\sim 700$ M dwarfs within 150 pc.

\begin{figure}
    \centering
    \includegraphics[width=\linewidth]{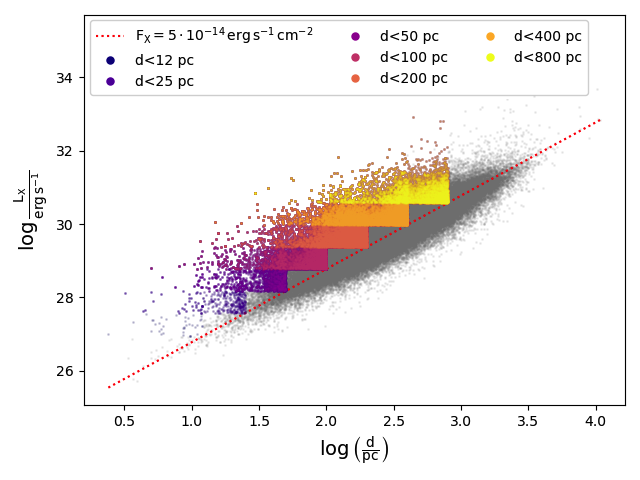}
    \caption{X-ray luminosity-distance phase space of the distance-limited samples (colors) and of the full HamStar catalog (grey points).}
    \label{fig:plot_Lx_vs_d}
\end{figure}

Fig.~\ref{fig:plot_Lx_vs_d} shows these subsamples (purple to yellow colors, consistent throughout all the following figures in this work) plotted on top of the whole HamStar sample (grey points). 
The stars matching our selection criteria show a homogeneous distribution within each 3D spherical volume with minor concentrations in the location of neighboring clusters (see Appendix~\ref{appendix:3D_spatial_distribution}).

\begin{figure}
    \centering
    \includegraphics[width=\linewidth]{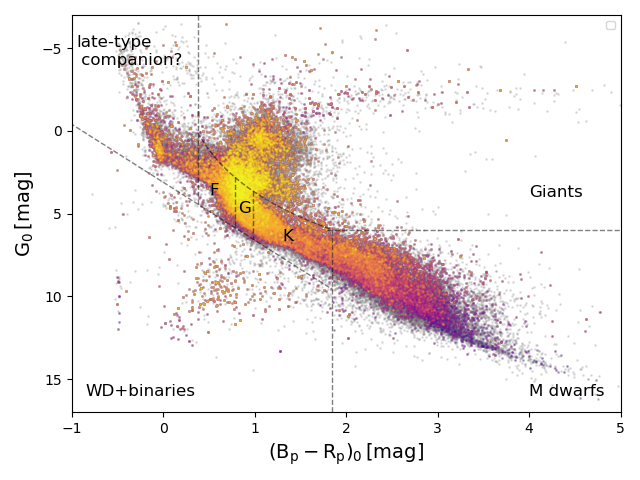}
    \caption{\textit{Gaia} DR3 color-magnitude diagram of the HamStar catalog (grey points). The colors correspond to Fig.~\ref{fig:plot_Lx_vs_d}.}
    \label{fig:plot_HR_HamStar}
\end{figure}

In order to identify the spectral type of the stars in our sample, we build the color-magnitude diagram $(G_0;\, (B_p-R_p)_0)$ shown in Fig.~\ref{fig:plot_HR_HamStar}. The subscript ${}_0$ indicates that the values have been corrected for extinction/reddening, when available in the \textit{Gaia} DR3 database ($74\%$ of our sample). When the GSP-Phot extinction values were not available ($26\%$ of our sample) we did not compute the correction and use the values as originally reported in HamStar. 
We define the main-sequence spectral types F, G, K and M using the \textit{Gaia} corrected colors, according to the online table of Eric Mamajek\footnote{"A Modern Mean Dwarf Stellar Color and Effective Temperature Sequence" - http://www.pas.rochester.edu/
/\~emamajek/EEM\_dwarf\_UBVIJHK\_colors\_Teff.txt - version 2022.04.16.}.
We assign all giant and supergiant stars to one group regardless of their color. The threshold line between main-sequence stars and giants is defined here as $G_0 \equiv 3.7 \cdot (1+\log_{10}(B_p-R_p)_0)$ for $(B_p-R_p)_0<1.84$, and $G_0 < 6$ for $(B_p-R_p)_0>1.84$. While not based on a physical motivation, this definition effectively separates the bulk of FGK main-sequence stars from the Giant Branch at similar $(B_p-R_p)_0$ \textit{Gaia} colors. 
A similar approach is taken to separate the White dwarfs (WD) and spectral type earlier than F.
The latter, in particular spectral types A and B, are usually characterized by radiative envelopes, where magnetic dynamo action is suppressed due to the lack of convection in the stellar envelope. The X-ray emission from these systems is thus most likely associated to an unresolved companion star \citep[e.g.][]{2006A&A...452.1001S}. We thus name this set as "late-type companion?" throughout the work. While giant stars are X-ray emitters, they are characterized in general by a slower rotation in their late evolutionary phase, and thus a likely less efficient dynamo than for main-sequence stars. Therefore, some of their X-ray flux may also be produced by a companion star. While a detailed characterization of each one of these sources is beyond the scope of this work, we rely on the HamStar identification of the Gaia counterpart as giant stars.
The WD subsample also includes binary systems with a main-sequence companion star. The sample is defined by $G_0 > 3.5 \cdot (B_p-R_p)_0 + 3.1$ for $(B_p-R_p)_0<1.84$ (i.e. earlier colors than M dwarfs). The division of the subsamples is illustrated in Fig.~\ref{fig:plot_HR_HamStar} with dashed lines.
Based on the large statistics available thanks to the large volumes, we tested that different definitions for the thresholds close to the one provided above do not produce significant changes in the derived XLF.

We also note that the median of the relative correction to the color is only $10-12\%$ across all spectral types. Therefore, the lack of extinction values on $26\%$ of our sample does not introduce a significant bias in the separation of the spectral types.

\section{Results}

\begin{figure}
    \centering
    \includegraphics[width=\linewidth]{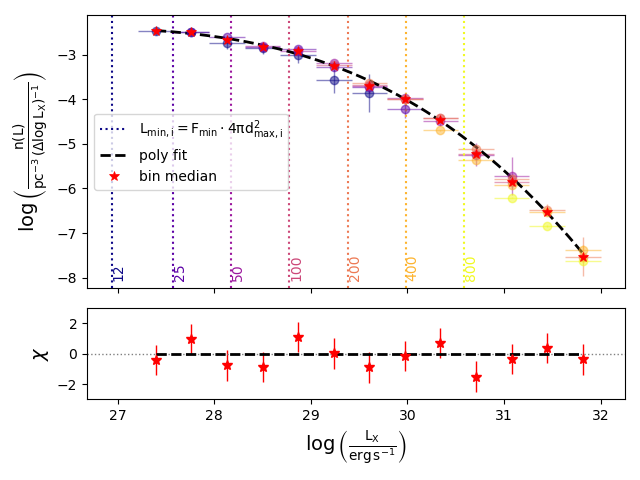}
    \caption{X-ray luminosity function of the stars in our sample (all). The colors of the vertical dotted lines match the ones of Fig.~\ref{fig:plot_Lx_vs_d} and show the minimum luminosity corresponding to the maximum distance in pc (label) of each subset.}
    \label{fig:plot_XLF_all_med_cut}
\end{figure}

\begin{figure*}
    \centering
    \includegraphics[width=\textwidth]{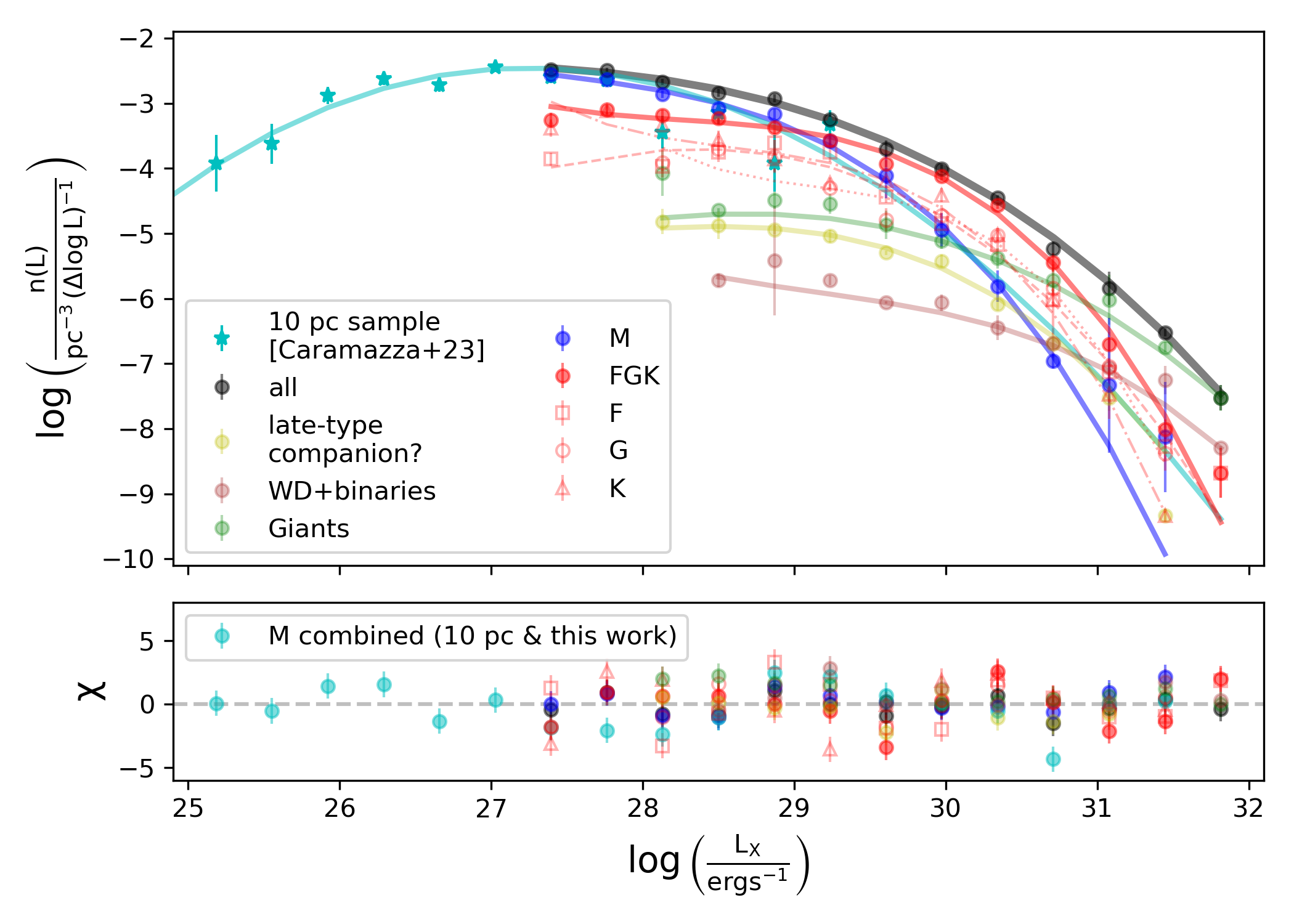}
    \caption{The X-ray luminosity function of stars in different parts of the color-magnitude diagram. The different colors show the contribution of the different stellar groups to the total XLF, as labeled.}
    \label{fig:plot_fL}
\end{figure*}

In Fig.~\ref{fig:plot_XLF_all_med_cut} we show the XLF derived with our method including all the stars, regardless of their group in the color-magnitude diagram. The XLF of each subsample defined by a given $\rm L_{min,i}$ is shown with its corresponding color. The XLF data points of adjacent subsamples (i.e. different $\rm L_{min,i}$) overlap within each of the shared luminosity bins. Smaller volumes in fact miss rare high-luminosity stars, while larger volumes miss low luminosity bins due to higher $\rm L_{min,i}$ threshold. The red star symbols show the median value in each luminosity bin. Altogether, they produce a smooth XLF extended over the $\rm L_X\in [2.5\times 10^{27}; 7.5\times{10^{31}}]\, erg\, s^{-1}$ range. 

With the same procedure, we computed the XLF for all the groups defined in Fig.~\ref{fig:plot_HR_HamStar}. The result is shown in Fig.~\ref{fig:plot_fL}. 
The red star symbols of Fig.~\ref{fig:plot_XLF_all_med_cut}, corresponding to the XLF values for all groups combined, are shown as the black points in Fig.~\ref{fig:plot_fL}.
We also derived the XLF from the 10 pc sample of \citealt[][]{2023A&A...676A..14C} (cyan star markers). Their volume-limited sample involves dedicated XMM-Newton follow-up. Therefore, it extends to lower X-ray luminosities with respect to the eROSITA-based HamStar catalog. In their work, most of the M dwarf stars within the closest 10 pc have been detected in X-rays (144 on 149 stars).
The XLF data points derived in this work for the M dwarf sample (blue dots) overlap with the XLF derived from the 10 pc sample in the higher $\rm L_X$ bins, confirming that our XLF is in fact a good approximation to a volume-complete result within the sensitivity range of eROSITA. 
We also note that the highest luminosity bins of the 10 pc sample have relatively large statistical uncertainty and scatter compared to the same bins of our M dwarf sample (blue points). This statistical improvement is in fact due to the approach of separating volume-limited samples of increasing luminosity and distance, generally providing larger samples.

We focus our results and discussion on the XLFs derived for the main-sequence spectral types F,G,K and M, while reporting also the results for the other stellar groups defined by our color-magnitude diagram for completeness. We also note that the spectral type assigned to sources labeled as "late-type companion?" may have more contamination from possible companion stars than those classified as F, G, K, or M types. However, as their X-ray emission still meets our flux–distance selection criteria, we include them in the computation of the overall XLF.
An additional test on true binary systems both resolved by \textit{Gaia} \citep{2021MNRAS.506.2269E} and eROSITA (separation $>10$") shows an XLF consistent with the one derived from the whole sample within $5\%$ (see Appendix~\ref{appendix:multiplicity} for details).

We find that M dwarfs dominate the low-luminosity end of the XLF $\rm L_X<10^{29}\, erg\, s^{-1}$. F, G and K type stars dominate in the $\rm 10^{29}-5\times 10^{30}\, erg\, s^{-1}$ luminosity range,  while at even higher luminosities giant and supergiant stars are the most frequent X-ray emitting stellar group (with their XLF poorly constrained below $\rm 10^{29}\, erg\, s^{-1}$).

\begin{table*}[]
    \caption{ Coefficients of the polynomial fit ${\rm XLF_{SpT}} = \Sigma_n\, a_n (\log L_X)^n$. 
     }
    \tiny
    \centering
    \begin{tabular}{lllll }
        Group & $a_3$ & $a_2$ & $a_1$ & $a_0$ \\
        \hline
        all & -2.64455e-02 & 2.06830e+00 & -5.39290e+01 & 4.66404e+02 \\
        M (this work)& -9.66074e-02 & 7.95632e+00 &-2.18718e+02 & 2.00432e+03 \\
        M combined$^\dagger$ & 3.28675e-03 & -6.14826e-01 & 2.61803e+01 & -3.25829e+02 \\
        FGK & -1.25846e-01 & 1.06682e+01 &-3.01598e+02 & 2.84023e+03 \\
        F & -4.17879e-02 & 3.22189e+00 &-8.19084e+01 & 6.81053e+02 \\
        G & -2.87207e-01 & 2.51049e+01 &-7.31781e+02 & 7.10888e+03 \\
        K & -2.08187e-01 & 1.78631e+01 &-5.11186e+02 & 4.87519e+03 \\
        giant & -1.92483e-02 & 1.42941e+00 &-3.44937e+01 & 2.62910e+02 \\
        late-type companions? & -6.81334e-02 & 5.6269e+00 & -1.54705e+02 & 1.41098e+03 \\
        WD + binaries & -7.91893e-02 & 6.92262e+00 & -2.02051e+02 & 1.96305e+03 \\
                        
    \end{tabular}
    \tablefoot{$\dagger$ joint fit of the 10 pc M dwarf sample (cyan star symbols) and the M dwarf group defined in this work (blue points).}
    \label{tab:poly_fit}
\end{table*}

We performed polynomial fits for each of the XLFs. The coefficients are reported in Table~\ref{tab:poly_fit}. We fixed the number of coefficients to 4 as we found it is the lowest degree to provide a reasonable fit for all stellar groups. The fitted polynomials are shown in Fig.~\ref{fig:plot_fL} as lines of colors corresponding to the data points of the different stellar groups.

\section{Discussion}

The knowledge on the low-luminosity end of the XLF is restricted towards the closest regions of the Solar neighborhood due to the instrumental flux limits. A complete census of all stellar X-ray emitters even in the smallest volume around
the Sun requires dedicated XMM-\textit{Newton} observations. this is achieved by the 10 pc survey described by \citet{2023A&A...676A..14C}. We find that the XLF derived in our work from eRASS1 matches the one derived from the volume-complete 10 pc M dwarf 10 pc sample in the common range of luminosities analyzed.
We note that \citealt{2023A&A...676A..14C} comprises only M dwarf stars of subtypes M0-M4. However, we find that subtypes M0-M4 account to $>96\%$ of the stars in the eRASS1 M dwarf sample, making a further selection for the comparison unnecessary.

Fig.~\ref{fig:plot_fL} shows that the high-luminosity tail of the XLF is missing from the 10 pc sample, as small volumes necessarily lack information on rare objects, such as those at the high-luminosity tail of the XLF. This implies that the average X-ray luminosity $\rm \langle L_X \rangle$ value from the 10 pc sample is biased low with respect to the overall galactic M dwarf population. 
We define the bias as $\rm 1 - C_{10 pc}/C_{M}$, where the latter element is the ratio between the integral $\rm C\equiv \int_{L_0} n/\Delta \log L'\, dL'$ derived from the 10 pc sample and the one computed in this work for M dwarfs, and $\rm L_0=10^{27}\, erg\, s^{-1}$, the sensitivity limit of eRASS1 for M dwarfs according to the HamStar catalog of \citet{2024A&A...684A.121F}. We obtain a bias of $2\%$. Given that our lower integration boundary, $\rm L_0$, is larger than the minimum luminosity reached by the 10 pc sample, the value of $2\%$ is actually an upper limit to the true bias. By joining the M dwarfs XLF derived in this work with the one derived for the 10 pc sample, and setting $\rm L_0=10^{24}\, erg\, s^{-1}$, we obtain a bias of only $0.3\%$.

\begin{figure}
    \centering
    \includegraphics[width=\linewidth]{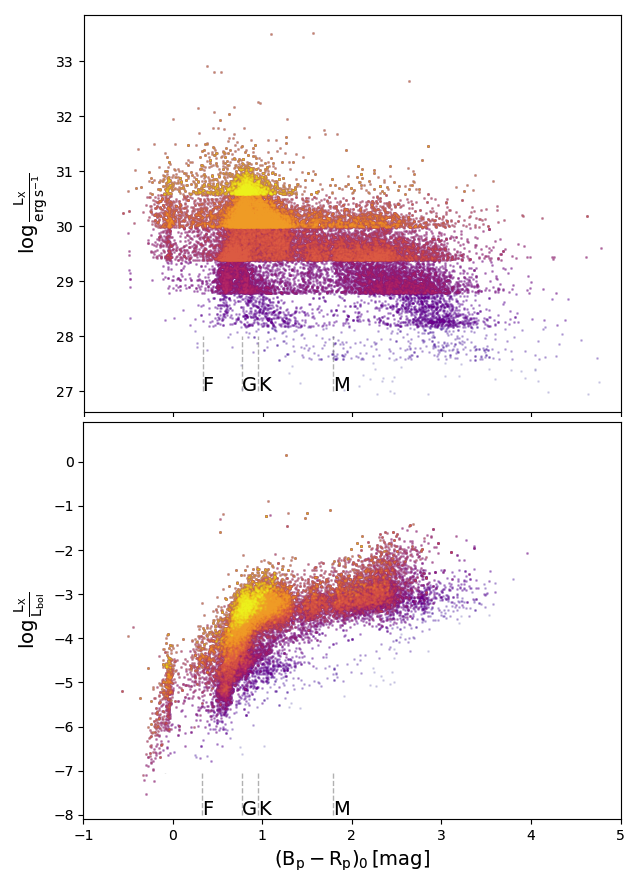}
    \caption{\textit{Upper panel:} Luminosity-color phase space of all the stars in our sample (apart from the giant group, for these see Appendix~\ref{appendix:giants}). The colors correspond to Fig.~\ref{fig:plot_Lx_vs_d}.
    \textit{Lower panel:} X-ray fractional luminosity versus Gaia color.}
    \label{fig:plot_Lx_vs_BpRp}
\end{figure}

In order to understand which luminosities are represented in which volumes for a given spectral type, in Fig.~\ref{fig:plot_Lx_vs_BpRp} (upper panel) we plot the X-ray luminosity $\rm L_X$ versus the \textit{Gaia} color $\rm (B_p-R_p)$ (extinction-corrected, when available) for all the stars in our sample (i.e. included late-type companions and WD+binaries groups), left aside the group of giants (see Appendix~\ref{appendix:giants} for the giant stars). In general, the number of stars increases in larger volumes until a certain value for the maximum distance (e.g. $\rm d_{max,i}=800\, pc$ for main-sequence stars F, G and K) where the corresponding luminosity threshold (e.g. $\rm L_{min,i}=5.4\times10^{30}\, erg\, s^{-1}$ for $\rm d_{max,i}=800\, pc$) is large enough to strongly limit the number of available sources. Almost no M dwarfs are in fact selected within the largest volumes. This fact is consistent with the general maximum luminosity of M dwarfs of a few times $\rm 10^{30}\, erg\, s^{-1}$ \citep{2020A&A...638A..20M, 2022A&A...661A..29M}. F, G and K stars reach larger X-ray luminosities instead and can be thus detected from larger distances.

In the lower panel of Fig.~\ref{fig:plot_Lx_vs_BpRp} we show a similar plot that considers the level of X-ray activity defined as $\rm L_X/L_{bol}$ (see Appendix~\ref{appendix:L_bol} for details on the bolometric luminosity $\rm L_{bol}$). The range in X-ray activity spans about 3 dex for each group (FGKM), but for
successively higher values. This plot clearly shows that our definition of the subsamples necessarily introduces a selection bias with respect to the X-ray activity. In fact, the continuous decrease of $\rm L_{bol}$ towards later spectral types limits the $\rm L_X/L_{bol}$ value of M dwarfs to larger values, as shown by the positive trend in the lower panel of Fig.~\ref{fig:plot_Lx_vs_BpRp}. This selection bias prevents an unbiased assessment of the distribution of $\rm L_X/L_{bol}$ of our sample. While its correct computation may be obtained by a forward modeling as a function of the other parameters ($\rm L_X$, spectral type), this approach is still limited by available volume-limited samples \citep{2025A&A...694A..93Z}. A forward modeling of the $\rm L_X/L_{bol}$ distribution thus necessarily goes well beyond the data-driven approach of this work.

An additional potential source of bias in the derived stellar density per luminosity bin $\rm n_{d\log L}$ may be introduced by a 3D disk-like spatial distribution of sources, compared to an homogeneous distribution. In fact, according to the definition $\rm n_{d\log L}\equiv N_{d\log L}/(4/3\pi d_{max}^3)$, the source distribution is assumed to be homogeneous over a spherical volume. The bias becomes important for $\rm d_{max}\gg z_h$ where $z_h$ is the disk scale-height. The thin and thick disk of the Milky Way have values of $\simeq300$ and 900 pc respectively \citep{2008ApJ...673..864J, 2017MNRAS.465...76M}. The bias is thus only potentially affecting our largest volume-luminosity bin, where M dwarfs are in practice not represented. This affects only a single data point in our derivation of the XLF (see Appendix~\ref{appendix:volume_bias}). Deeper surveys ($\rm d_{max}\geq 1$ kpc) should however take this bias in consideration.

\section{Conclusion}

Using the first eROSITA all-sky survey data and its stellar identification from the HamStar catalog \citep{2024A&A...684A.121F}, we have computed the XLF of stars, per stellar group, probing distances up to 800 pc from the Sun.
We find that for the eRASS1 sensitivity limit $\rm F_{lim}=5\times10^{14}\, erg\, s^{-1}\, cm^{-2}$ M dwarfs, FGK types stars and giants dominate the XLF above $\rm L_X \sim 10^{27},\, 10^{29}\, and\, 5\times 10^{30}\, erg\, s^{-1}$, respectively.
The previously known XLF of M dwarfs probes the closest 10 pc to the Sun \citep{2023A&A...676A..14C, 2025A&A...694A..93Z}. The comparison with our sample shows that the volume-complete (10 pc) approach probes the lowest X-ray luminosities for this spectral type, which are only accessible in the smallest volumes, but it misses the rare high-luminosity stars. The XLF derived in this work is able to probe both the high-luminosity end and to explore the XLF over the largest volume to date for sufficiently high X-ray luminosities, both for M dwarfs and other spectral types.
We provide polynomial fits for the XLF of all stellar groups, including main-sequence stars (F, G, K and M), giants and WD+binaries. The results from our polynomial fits to the XLF can be used to model the X-ray stellar contribution to the unresolved soft X-ray emission of Milky Way-like galaxies.
Improved knowledge on both ends of the XLF (and of the X-ray activity) can only come from future all-sky surveys reaching lower X-ray fluxes than eROSITA or from complete surveys that are however feasible only for very small sky volumes.

\begin{acknowledgements}
We acknowledge financial support from the European Research Council (ERC) under the European Union’s Horizon 2020 research and innovation program Hot- Milk (grant agreement No 865637). GP acknowledges support from Bando per il Finanziamento della Ricerca Fondamentale 2022 dell’Istituto Nazionale di Astrofisica (INAF): GO Large program and from the Framework per l’Attrazione e il Rafforzamento delle Eccellenze (FARE) per la ricerca in Italia (R20L5S39T9). E.M. was supported by Deutsche Forschungsgemeinschaft under grant STE 1068/8-1. We thank the anonymous reviewer for helping to improve the manuscript.
\end{acknowledgements}

\bibliography{aa} 

\begin{thebibliography}{30}
\expandafter\ifx\csname natexlab\endcsname\relax\def\natexlab#1{#1}\fi

\bibitem[{{Bailer-Jones}(2011)}]{2011MNRAS.411..435B}
{Bailer-Jones}, C.~A.~L. 2011, \mnras, 411, 435

\bibitem[{{Belczynski} {et~al.}(2004){Belczynski}, {Kalogera}, {Zezas}, \& {Fabbiano}}]{2004ApJ...601L.147B}
{Belczynski}, K., {Kalogera}, V., {Zezas}, A., \& {Fabbiano}, G. 2004, \apjl, 601, L147

\bibitem[{{Belczynski} \& {Taam}(2004)}]{2004ApJ...603..690B}
{Belczynski}, K. \& {Taam}, R.~E. 2004, \apj, 603, 690

\bibitem[{{Caramazza} {et~al.}(2023){Caramazza}, {Stelzer}, {Magaudda}, {Raetz}, {G{\"u}del}, {Orlando}, \& {Poppenh{\"a}ger}}]{2023A&A...676A..14C}
{Caramazza}, M., {Stelzer}, B., {Magaudda}, E., {et~al.} 2023, \aap, 676, A14

\bibitem[{{Creevey} {et~al.}(2023){Creevey}, {Sordo}, {Pailler}, {Fr{\'e}mat}, {Heiter}, {Th{\'e}venin}, {Andrae}, {Fouesneau}, {Lobel}, {Bailer-Jones}, {Garabato}, {Bellas-Velidis}, {Brugaletta}, {Lorca}, {Ordenovic}, {Palicio}, {Sarro}, {Delchambre}, {Drimmel}, {Rybizki}, {Torralba Elipe}, {Korn}, {Recio-Blanco}, {Schultheis}, {De Angeli}, {Montegriffo}, {Abreu Aramburu}, {Accart}, {{\'A}lvarez}, {Bakker}, {Brouillet}, {Burlacu}, {Carballo}, {Casamiquela}, {Chiavassa}, {Contursi}, {Cooper}, {Dafonte}, {Dapergolas}, {de Laverny}, {Dharmawardena}, {Edvardsson}, {Le Fustec}, {Garc{\'\i}a-Lario}, {Garc{\'\i}a-Torres}, {Gomez}, {Gonz{\'a}lez-Santamar{\'\i}a}, {Hatzidimitriou}, {Jean-Antoine Piccolo}, {Kontiza}, {Kordopatis}, {Lanzafame}, {Lebreton}, {Licata}, {Lindstr{\o}m}, {Livanou}, {Magdaleno Romeo}, {Manteiga}, {Marocco}, {Marshall}, {Mary}, {Nicolas}, {Pallas-Quintela}, {Panem}, {Pichon}, {Poggio}, {Riclet}, {Robin}, {Santove{\~n}a}, {Silvelo}, {Slezak}, {Smart}, {Soubiran}, {S{\"u}veges}, {Ulla},
  {Utrilla}, {Vallenari}, {Zhao}, {Zorec}, {Barrado}, {Bijaoui}, {Bouret}, {Blomme}, {Brott}, {Cassisi}, {Kochukhov}, {Martayan}, {Shulyak}, \& {Silvester}}]{2023A&A...674A..26C}
{Creevey}, O.~L., {Sordo}, R., {Pailler}, F., {et~al.} 2023, \aap, 674, A26

\bibitem[{{Donati} {et~al.}(2008){Donati}, {Morin}, {Petit}, {Delfosse}, {Forveille}, {Auri{\`e}re}, {Cabanac}, {Dintrans}, {Fares}, {Gastine}, {Jardine}, {Ligni{\`e}res}, {Paletou}, {Ramirez Velez}, \& {Th{\'e}ado}}]{2008MNRAS.390..545D}
{Donati}, J.~F., {Morin}, J., {Petit}, P., {et~al.} 2008, \mnras, 390, 545

\bibitem[{{El-Badry} {et~al.}(2021){El-Badry}, {Rix}, \& {Heintz}}]{2021MNRAS.506.2269E}
{El-Badry}, K., {Rix}, H.-W., \& {Heintz}, T.~M. 2021, \mnras, 506, 2269

\bibitem[{{Freund} {et~al.}(2024){Freund}, {Czesla}, {Predehl}, {Robrade}, {Salvato}, {Schneider}, {Starck}, {Wolf}, \& {Schmitt}}]{2024A&A...684A.121F}
{Freund}, S., {Czesla}, S., {Predehl}, P., {et~al.} 2024, \aap, 684, A121

\bibitem[{{Gaia Collaboration} {et~al.}(2021){Gaia Collaboration}, {Brown}, {Vallenari}, {Prusti}, {de Bruijne}, {Babusiaux}, {Biermann}, {Creevey}, {Evans}, {Eyer}, {Hutton}, {Jansen}, {Jordi}, {Klioner}, {Lammers}, {Lindegren}, {Luri}, {Mignard}, {Panem}, {Pourbaix}, {Randich}, {Sartoretti}, {Soubiran}, {Walton}, {Arenou}, {Bailer-Jones}, {Bastian}, {Cropper}, {Drimmel}, {Katz}, {Lattanzi}, {van Leeuwen}, {Bakker}, {Cacciari}, {Casta{\~n}eda}, {De Angeli}, {Ducourant}, {Fabricius}, {Fouesneau}, {Fr{\'e}mat}, {Guerra}, {Guerrier}, {Guiraud}, {Jean-Antoine Piccolo}, {Masana}, {Messineo}, {Mowlavi}, {Nicolas}, {Nienartowicz}, {Pailler}, {Panuzzo}, {Riclet}, {Roux}, {Seabroke}, {Sordo}, {Tanga}, {Th{\'e}venin}, {Gracia-Abril}, {Portell}, {Teyssier}, {Altmann}, {Andrae}, {Bellas-Velidis}, {Benson}, {Berthier}, {Blomme}, {Brugaletta}, {Burgess}, {Busso}, {Carry}, {Cellino}, {Cheek}, {Clementini}, {Damerdji}, {Davidson}, {Delchambre}, {Dell'Oro}, {Fern{\'a}ndez-Hern{\'a}ndez}, {Galluccio}, {Garc{\'\i}a-Lario},
  {Garcia-Reinaldos}, {Gonz{\'a}lez-N{\'u}{\~n}ez}, {Gosset}, {Haigron}, {Halbwachs}, {Hambly}, {Harrison}, {Hatzidimitriou}, {Heiter}, {Hern{\'a}ndez}, {Hestroffer}, {Hodgkin}, {Holl}, {Jan{\ss}en}, {Jevardat de Fombelle}, {Jordan}, {Krone-Martins}, {Lanzafame}, {L{\"o}ffler}, {Lorca}, {Manteiga}, {Marchal}, {Marrese}, {Moitinho}, {Mora}, {Muinonen}, {Osborne}, {Pancino}, {Pauwels}, {Petit}, {Recio-Blanco}, {Richards}, {Riello}, {Rimoldini}, {Robin}, {Roegiers}, {Rybizki}, {Sarro}, {Siopis}, {Smith}, {Sozzetti}, {Ulla}, {Utrilla}, {van Leeuwen}, {van Reeven}, {Abbas}, {Abreu Aramburu}, {Accart}, {Aerts}, {Aguado}, {Ajaj}, {Altavilla}, {{\'A}lvarez}, {{\'A}lvarez Cid-Fuentes}, {Alves}, {Anderson}, {Anglada Varela}, {Antoja}, {Audard}, {Baines}, {Baker}, {Balaguer-N{\'u}{\~n}ez}, {Balbinot}, {Balog}, {Barache}, {Barbato}, {Barros}, {Barstow}, {Bartolom{\'e}}, {Bassilana}, {Bauchet}, {Baudesson-Stella}, {Becciani}, {Bellazzini}, {Bernet}, {Bertone}, {Bianchi}, {Blanco-Cuaresma}, {Boch}, {Bombrun}, {Bossini},
  {Bouquillon}, {Bragaglia}, {Bramante}, {Breedt}, {Bressan}, {Brouillet}, {Bucciarelli}, {Burlacu}, {Busonero}, {Butkevich}, {Buzzi}, {Caffau}, {Cancelliere}, {C{\'a}novas}, {Cantat-Gaudin}, {Carballo}, {Carlucci}, {Carnerero}, {Carrasco}, {Casamiquela}, {Castellani}, {Castro-Ginard}, {Castro Sampol}, {Chaoul}, {Charlot}, {Chemin}, {Chiavassa}, {Cioni}, {Comoretto}, {Cooper}, {Cornez}, {Cowell}, {Crifo}, {Crosta}, {Crowley}, {Dafonte}, {Dapergolas}, {David}, \& {David}}]{2021A&A...649A...1G}
{Gaia Collaboration}, {Brown}, A.~G.~A., {Vallenari}, A., {et~al.} 2021, \aap, 649, A1

\bibitem[{{G{\"u}del}(2004)}]{2004A&ARv..12...71G}
{G{\"u}del}, M. 2004, \aapr, 12, 71

\bibitem[{{Gustafsson} {et~al.}(2008){Gustafsson}, {Edvardsson}, {Eriksson}, {J{\o}rgensen}, {Nordlund}, \& {Plez}}]{2008A&A...486..951G}
{Gustafsson}, B., {Edvardsson}, B., {Eriksson}, K., {et~al.} 2008, \aap, 486, 951

\bibitem[{{Johnstone} {et~al.}(2021){Johnstone}, {Bartel}, \& {G{\"u}del}}]{2021A&A...649A..96J}
{Johnstone}, C.~P., {Bartel}, M., \& {G{\"u}del}, M. 2021, \aap, 649, A96

\bibitem[{{Johnstone} \& {G{\"u}del}(2015)}]{2015A&A...578A.129J}
{Johnstone}, C.~P. \& {G{\"u}del}, M. 2015, \aap, 578, A129

\bibitem[{{Juri{\'c}} {et~al.}(2008){Juri{\'c}}, {Ivezi{\'c}}, {Brooks}, {Lupton}, {Schlegel}, {Finkbeiner}, {Padmanabhan}, {Bond}, {Sesar}, {Rockosi}, {Knapp}, {Gunn}, {Sumi}, {Schneider}, {Barentine}, {Brewington}, {Brinkmann}, {Fukugita}, {Harvanek}, {Kleinman}, {Krzesinski}, {Long}, {Neilsen}, {Nitta}, {Snedden}, \& {York}}]{2008ApJ...673..864J}
{Juri{\'c}}, M., {Ivezi{\'c}}, {\v{Z}}., {Brooks}, A., {et~al.} 2008, \apj, 673, 864

\bibitem[{{Kiraga} \& {Stepien}(2007)}]{2007AcA....57..149K}
{Kiraga}, M. \& {Stepien}, K. 2007, \actaa, 57, 149

\bibitem[{{Magaudda} {et~al.}(2020){Magaudda}, {Stelzer}, {Covey}, {Raetz}, {Matt}, \& {Scholz}}]{2020A&A...638A..20M}
{Magaudda}, E., {Stelzer}, B., {Covey}, K.~R., {et~al.} 2020, \aap, 638, A20

\bibitem[{{Magaudda} {et~al.}(2022){Magaudda}, {Stelzer}, {Raetz}, {Klutsch}, {Salvato}, \& {Wolf}}]{2022A&A...661A..29M}
{Magaudda}, E., {Stelzer}, B., {Raetz}, S., {et~al.} 2022, \aap, 661, A29

\bibitem[{{McMillan}(2017)}]{2017MNRAS.465...76M}
{McMillan}, P.~J. 2017, \mnras, 465, 76

\bibitem[{{Merloni} {et~al.}(2024){Merloni}, {Lamer}, {Liu}, {Ramos-Ceja}, {Brunner}, {Bulbul}, {Dennerl}, {Doroshenko}, {Freyberg}, {Friedrich}, {Gatuzz}, {Georgakakis}, {Haberl}, {Igo}, {Kreykenbohm}, {Liu}, {Maitra}, {Malyali}, {Mayer}, {Nandra}, {Predehl}, {Robrade}, {Salvato}, {Sanders}, {Stewart}, {Tub{\'\i}n-Arenas}, {Weber}, {Wilms}, {Arcodia}, {Artis}, {Aschersleben}, {Avakyan}, {Aydar}, {Bahar}, {Balzer}, {Becker}, {Berger}, {Boller}, {Bornemann}, {Br{\"u}ggen}, {Brusa}, {Buchner}, {Burwitz}, {Camilloni}, {Clerc}, {Comparat}, {Coutinho}, {Czesla}, {Dannhauer}, {Dauner}, {Dauser}, {Dietl}, {Dolag}, {Dwelly}, {Egg}, {Ehl}, {Freund}, {Friedrich}, {Gaida}, {Garrel}, {Ghirardini}, {Gokus}, {Gr{\"u}nwald}, {Grandis}, {Grotova}, {Gruen}, {Gueguen}, {H{\"a}mmerich}, {Hamaus}, {Hasinger}, {Haubner}, {Homan}, {Ider Chitham}, {Joseph}, {Joyce}, {K{\"o}nig}, {Kaltenbrunner}, {Khokhriakova}, {Kink}, {Kirsch}, {Kluge}, {Knies}, {Krippendorf}, {Krumpe}, {Kurpas}, {Li}, {Liu}, {Locatelli}, {Lorenz}, {M{\"u}ller},
  {Magaudda}, {Mannes}, {McCall}, {Meidinger}, {Michailidis}, {Migkas}, {Mu{\~n}oz-Giraldo}, {Musiimenta}, {Nguyen-Dang}, {Ni}, {Olechowska}, {Ota}, {Pacaud}, {Pasini}, {Perinati}, {Pires}, {Pommranz}, {Ponti}, {Poppenhaeger}, {P{\"u}hlhofer}, {Rau}, {Reh}, {Reiprich}, {Roster}, {Saeedi}, {Santangelo}, {Sasaki}, {Schmitt}, {Schneider}, {Schrabback}, {Schuster}, {Schwope}, {Seppi}, {Serim}, {Shreeram}, {Sokolova-Lapa}, {Starck}, {Stelzer}, {Stierhof}, {Suleimanov}, {Tenzer}, {Traulsen}, {Tr{\"u}mper}, {Tsuge}, {Urrutia}, {Veronica}, {Waddell}, {Willer}, {Wolf}, {Yeung}, {Zainab}, {Zangrandi}, {Zhang}, {Zhang}, \& {Zheng}}]{2024A&A...682A..34M}
{Merloni}, A., {Lamer}, G., {Liu}, T., {et~al.} 2024, \aap, 682, A34

\bibitem[{{Miyaji} {et~al.}(2000){Miyaji}, {Hasinger}, \& {Schmidt}}]{2000A&A...353...25M}
{Miyaji}, T., {Hasinger}, G., \& {Schmidt}, M. 2000, \aap, 353, 25

\bibitem[{{Pizzocaro} {et~al.}(2019){Pizzocaro}, {Stelzer}, {Poretti}, {Raetz}, {Micela}, {Belfiore}, {Marelli}, {Salvetti}, \& {De Luca}}]{2019A&A...628A..41P}
{Pizzocaro}, D., {Stelzer}, B., {Poretti}, E., {et~al.} 2019, \aap, 628, A41

\bibitem[{{Predehl} {et~al.}(2021){Predehl}, {Andritschke}, {Arefiev}, {Babyshkin}, {Batanov}, {Becker}, {B{\"o}hringer}, {Bogomolov}, {Boller}, {Borm}, {Bornemann}, {Br{\"a}uninger}, {Br{\"u}ggen}, {Brunner}, {Brusa}, {Bulbul}, {Buntov}, {Burwitz}, {Burkert}, {Clerc}, {Churazov}, {Coutinho}, {Dauser}, {Dennerl}, {Doroshenko}, {Eder}, {Emberger}, {Eraerds}, {Finoguenov}, {Freyberg}, {Friedrich}, {Friedrich}, {F{\"u}rmetz}, {Georgakakis}, {Gilfanov}, {Granato}, {Grossberger}, {Gueguen}, {Gureev}, {Haberl}, {H{\"a}lker}, {Hartner}, {Hasinger}, {Huber}, {Ji}, {Kienlin}, {Kink}, {Korotkov}, {Kreykenbohm}, {Lamer}, {Lomakin}, {Lapshov}, {Liu}, {Maitra}, {Meidinger}, {Menz}, {Merloni}, {Mernik}, {Mican}, {Mohr}, {M{\"u}ller}, {Nandra}, {Nazarov}, {Pacaud}, {Pavlinsky}, {Perinati}, {Pfeffermann}, {Pietschner}, {Ramos-Ceja}, {Rau}, {Reiffers}, {Reiprich}, {Robrade}, {Salvato}, {Sanders}, {Santangelo}, {Sasaki}, {Scheuerle}, {Schmid}, {Schmitt}, {Schwope}, {Shirshakov}, {Steinmetz}, {Stewart}, {Str{\"u}der},
  {Sunyaev}, {Tenzer}, {Tiedemann}, {Tr{\"u}mper}, {Voron}, {Weber}, {Wilms}, \& {Yaroshenko}}]{Predehl2021}
{Predehl}, P., {Andritschke}, R., {Arefiev}, V., {et~al.} 2021, \aap, 647, A1

\bibitem[{{Preibisch} \& {Feigelson}(2005)}]{2005ApJS..160..390P}
{Preibisch}, T. \& {Feigelson}, E.~D. 2005, \apjs, 160, 390

\bibitem[{{Revnivtsev} {et~al.}(2009){Revnivtsev}, {Sazonov}, {Churazov}, {Forman}, {Vikhlinin}, \& {Sunyaev}}]{2009Natur.458.1142R}
{Revnivtsev}, M., {Sazonov}, S., {Churazov}, E., {et~al.} 2009, \nat, 458, 1142

\bibitem[{{Schmitt} \& {Liefke}(2004)}]{2004A&A...417..651S}
{Schmitt}, J.~H.~M.~M. \& {Liefke}, C. 2004, \aap, 417, 651

\bibitem[{{Stelzer} {et~al.}(2006){Stelzer}, {Hu{\'e}lamo}, {Micela}, \& {Hubrig}}]{2006A&A...452.1001S}
{Stelzer}, B., {Hu{\'e}lamo}, N., {Micela}, G., \& {Hubrig}, S. 2006, \aap, 452, 1001

\bibitem[{{Stelzer} {et~al.}(2022){Stelzer}, {Klutsch}, {Coffaro}, {Magaudda}, \& {Salvato}}]{2022A&A...661A..44S}
{Stelzer}, B., {Klutsch}, A., {Coffaro}, M., {Magaudda}, E., \& {Salvato}, M. 2022, \aap, 661, A44

\bibitem[{{Strong}(2007)}]{2007Ap&SS.309...35S}
{Strong}, A.~W. 2007, \apss, 309, 35

\bibitem[{{Wright} {et~al.}(2018){Wright}, {Newton}, {Williams}, {Drake}, \& {Yadav}}]{2018MNRAS.479.2351W}
{Wright}, N.~J., {Newton}, E.~R., {Williams}, P. K.~G., {Drake}, J.~J., \& {Yadav}, R.~K. 2018, \mnras, 479, 2351

\bibitem[{{Zhu} \& {Preibisch}(2025)}]{2025A&A...694A..93Z}
{Zhu}, E. \& {Preibisch}, T. 2025, \aap, 694, A93

\end{thebibliography}

\begin{appendix} 

\section{3D spatial distribution} \label{appendix:3D_spatial_distribution}

\begin{figure}
    \centering
    \includegraphics[width=\linewidth]{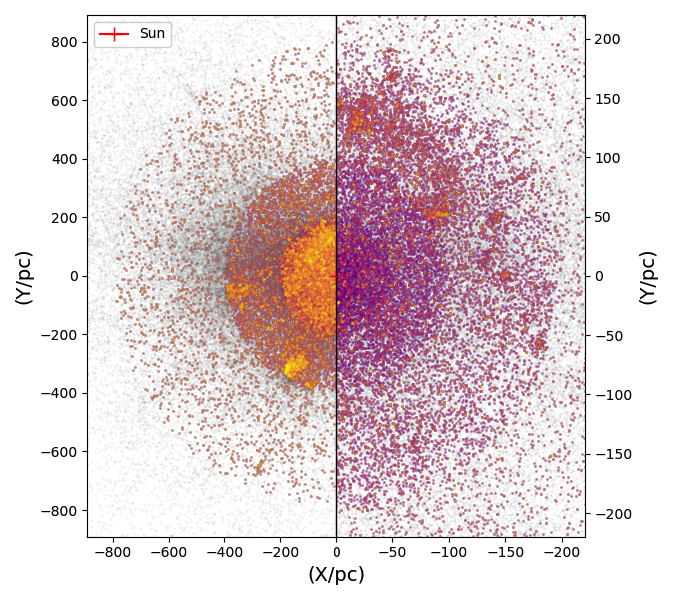}
    \caption{X-Y projection of the position of the HamStar catalog (grey points). The selection of the volume-limited subsamples is shown with colors corresponding to Fig.~\ref{fig:plot_Lx_vs_d}. The right half of the plot shows a zoom-in of the nearest 200 pc to the Sun with mirrored X coordinate.}
    \label{fig:plot_XY_200+800pc}
\end{figure}

\begin{figure}
    \centering
    \includegraphics[width=\linewidth]{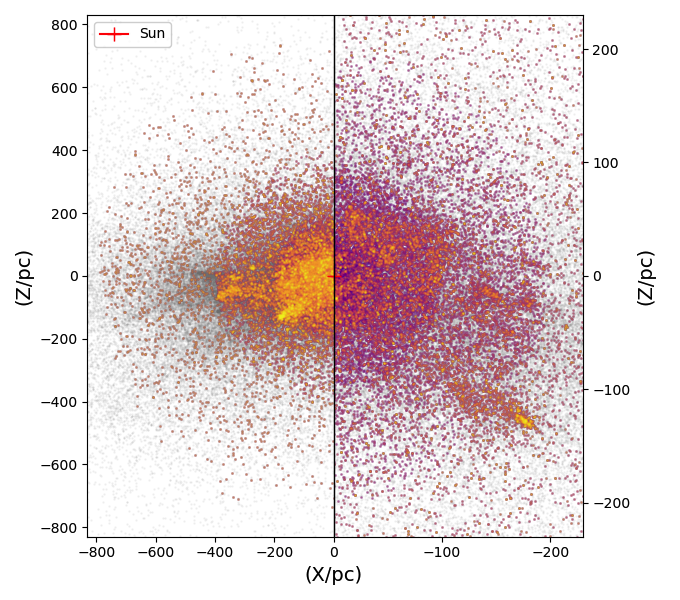}
    \caption{X-Z projection, same as Fig.~\ref{fig:plot_XY_200+800pc}.}
    \label{fig:plot_XZ_200+800pc}
\end{figure}

In Fig.~\ref{fig:plot_XY_200+800pc} and \ref{fig:plot_XZ_200+800pc} we plot the location of the stars in our sample, in 2D projections around the Sun as seen from far above (XY) and within (XZ) the Milky Way mid-plane, respectively. Given the broad range of scales in distances of the different subsamples, the right half of the plots zooms in the closest 200 pc to the Sun to highlight the position of the closest stars, i.e. as the HamStar sample only explores the Western Galactic sky based on the data available to the German eROSITA Consortium, the right-hand part of the Figures has a mirrored X-axis. It can be seen that the stars are mostly homogeneously distributed within each 3D spherical volume. It is also possible to identify clusters of stars. Given the relative size and population of these overdensities, we conclude that our selection is not strongly biased towards a particular direction in the sky.

\section{Multiplicity} \label{appendix:multiplicity}

Given the higher spatial resolution of the \textit{Gaia} DR3 data compared to eROSITA ($\rm \sim 1.5\, and\, 10$ arcsec, respectively), several optical counterparts for a given X-ray source are often found in the positional cross-match. This issue is already discussed in detail in the HamStar description \citep{2024A&A...684A.121F} for the interested reader.
While we do not repeat the cross-matches, we verify that our XLF is not influenced by the selection of the wrong
optical counterpart in binary systems. To this end, we consider a subsample of the one used in this work that includes all and only the stars recognized as part of a binary system with angular separation $\Delta\theta$ between the companion stars i.e. larger than the eROSITA resolution of 10 arcsec). 
This is possible thanks to the catalog of \textit{Gaia} binary systems \cite{2021MNRAS.506.2269E}. This catalog is based on the \textit{Gaia} DR3 sources and is cleaned from chance positional coincidence of stars \footnote{For all the \textit{Gaia} stars it is not possible to tell whether they hold or not a companion closer than the \textit{Gaia} resolution of 1.5 arcsec and we know all the binaries resolved by \textit{Gaia} but not by eROSITA ($1.5>\Delta\theta>10$ arcsec).}. By selecting $\Delta\theta>10$ arcsec, we obtain only stars that are known to have a companion, but the companion can be resolved with eROSITA. 

\begin{figure}
    \centering
    \includegraphics[width=\linewidth]{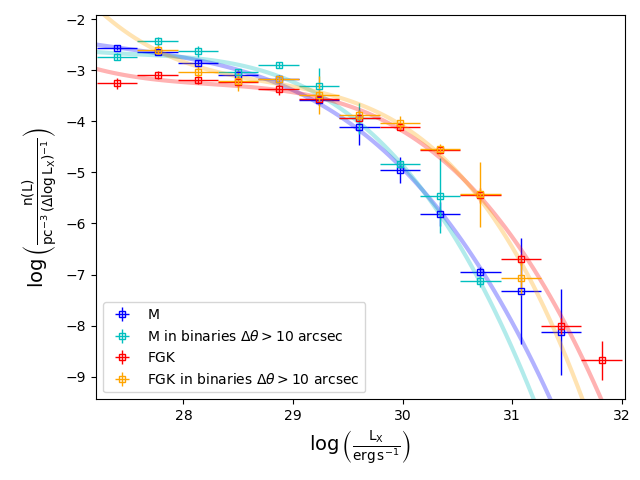}
    \caption{Comparison between the XLF for the subsample of members of binary systems with $\Delta\theta >10$ arcsec (red squares) and the XLF derived from the whole sample used in this work (black dots, equal to Fig.~\ref{fig:plot_fL}). }
    \label{fig:plot_XLF_binaryBias}
\end{figure}
From the cross-match with the catalog by \citet{2021MNRAS.506.2269E}, we find that $\sim 22\%$ of the whole sample used in this work is part of binary systems with $\Delta\theta>10$ arcsec. The XLFs built on this additional selection (binaries with $\Delta\theta>10$ arcsec, cyan and orange squares in Fig.~\ref{fig:plot_XLF_binaryBias} for the M and FGK groups, respectively) match the XLFs of the original group samples  (blue and red squares for the M and FGK groups, respectively) always within $5\%$ for all luminosity bins. This result shows that our derivation of the XLF is relatively robust against spurious cross-match due to multiplicity.

\section{Giant stars} \label{appendix:giants}

\begin{figure}
    \centering
    \includegraphics[width=0.9\linewidth]{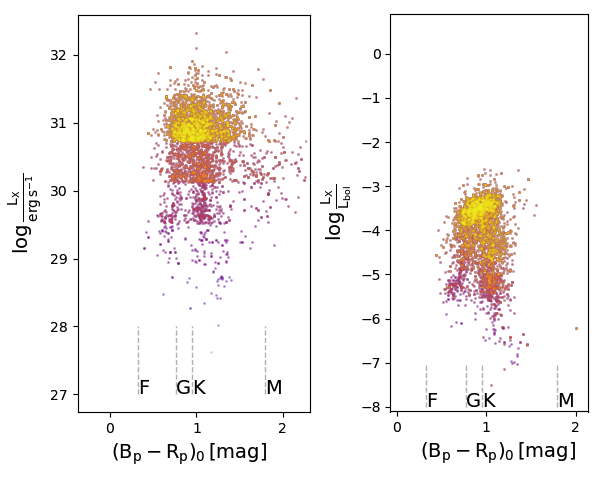}
    \caption{\textit{Left panel:} X-ray luminosity-color phase space of the distance subsamples for the group of the giant stars. \textit{Right panel:} X-ray activity-color for the same samples. The colors correspond to Fig.~\ref{fig:plot_Lx_vs_d}.}
    \label{fig:plot_Lx_vs_BpRp_giants}
\end{figure}

In Fig.~\ref{fig:plot_Lx_vs_BpRp_giants} we show equivalents of the upper and lower panels of Fig.~\ref{fig:plot_Lx_vs_BpRp} (left and right panel, respectively), for the giants group. We note that most of the giants show very high X-ray luminosity that are statistically more likely to be found in the largest volumes given the relatively low frequency. We note that most of the population of M-type giant stars may be missing, although they are the majority of all giants according to stellar models.

\section{Exponential vs. homogeneous height distribution} \label{appendix:volume_bias}

We verified numerically that for a known XLF, the one derived applying our selection criteria is not affected for the values of $\rm d_{max}$ reached in this work whether assuming a homogeneous or disk-like distribution of sources. 
To this end, for simplicity we assume a power-law XLF with low- and high- luminosity cut-offs. We generate X-ray sources at random locations (X, Y, Z) around the Sun where X and Y lay on the Galactic midplane and Z is the height above the plane. We draw X, Y and Z from uniform probability distribution functions. The X-ray luminosity of each source is randomly drawn from the assumed power-law XLF. We then apply our selection criteria of $\rm L_X>L_{X,min}\, \&\, d<d_{max}(L_{X,min})$ to the mock sample. The number of simulated sources is chosen in order to match (after selection) the number of sources of the observed sample. We check that the XLF derived from this mock sample reproduces the assumed one in all luminosity bins within the statistical uncertainties.

\begin{figure}
    \centering
    \includegraphics[width=\linewidth]{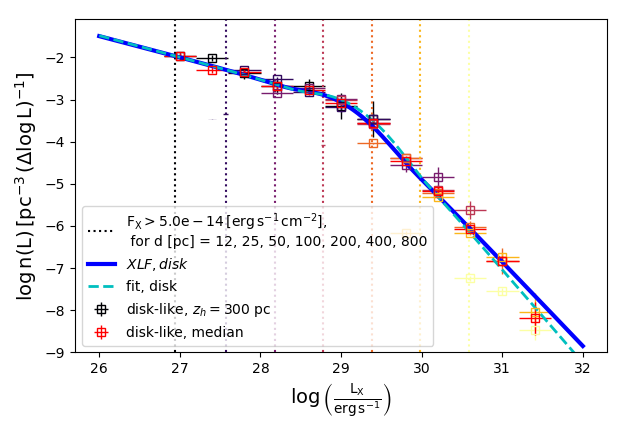}
    \caption{Simulated XLF from a disk-like distribution of sources.}
    \label{fig:plot_volumeBias_median_zh300pc}
\end{figure}

We then repeated the simulation by only varying the probability distribution of the Z coordinate $\rm PDF(Z)$ in order to test a different distribution of the stars with height from the Galactic plane. This time we draw the Z coordinate from the function $\rm PDF(Z)\propto e^{-Z/Z_h}$. The X and Y coordinates and the X-ray luminosities are drawn again, in the same way as in the simulation for homogeneous spherical distribution, outlined above. The result is shown in Fig.~\ref{fig:plot_volumeBias_median_zh300pc}. The computed XLF (the red squares are the median values in each luminosity bin) are fitted with a power-law model (cyan dashed line) which is in good agreement with the assumed XLF (blue solid line) within the statistical uncertainties for all values of $\rm Z_h \leq 400$ pc. This is due to the fact that for $\rm Z< Z_h,\, PDF(z) \longrightarrow const$, so that we fall again in the homogeneous scenario. The values shown by the yellow squares are high-luminosity sources in the largest volume. They are possibly biased low due to the thin disk geometrical effect, as shown by the slightly lower fit values (cyan dashed line) compared to the assumed XLF (blue solid line) in these bins. However, the real case scenario is composed by the sum of a thin ($\rm Z_h\simeq 300$ pc) and a thick disk ($\rm Z_h\simeq 900$ pc). The additional presence of a thick disk thus smoothens this biases further compared to our simulation. The total bias is thus negligible for the distance values considered in this work.

\section{Bolometric luminosity $\rm L_{bol}$} \label{appendix:L_bol}

The bolometric luminosity $\rm L_{bol}$ is computed through the standard set of equations

\begin{equation}
    \rm M_{bol} - M_{bol,\odot} = -2.5 \log (L_{bol}/L_{bol,\odot})
\end{equation}
\begin{equation}
        \rm M_{bol} = G_0 + BC_G
\end{equation}

where $\rm M_{bol,\odot}=4.66$ is the absolute bolometric magnitude of the Sun, as calibrated by Gaia  \citep{2023A&A...674A..26C} to estimate the bolometric correction $\rm BC_G$ to the G band extinction-corrected absolute magnitude $\rm G_0$. 
The bolometric correction $\rm BC_G$ is tabulated in the \texttt{bc\_flame} parameter columns of the Gaia DR3 \texttt{astrophysical\_parameter} table 
\footnote{\url{https://gea.esac.esa.int/archive/documentation/GDR3/Gaia_archive/chap_datamodel/sec_dm_astrophysical_parameter_tables/ssec_dm_astrophysical_parameters.html}}. 
The \texttt{bc\_flame} value is a function of effective temperature, surface gravity, and metallicity, and has been derived from MARCS synthetic stellar spectra library \citep{2008A&A...486..951G} applied to the GSP-Photometry results on the Gaia DR3 sources. We note that these are the same Gaia DR3 entries for which extinction correction information was computed.
The bolometric luminosity of the Sun is $\rm L_{bol,\odot}=3.828\times 10^{33}\, erg\, s^{-1}$.

\end{appendix}

\end{document}